\documentclass[prx,aps,superscriptaddress,footinbib,twocolumn]{revtex4-2}
\usepackage[english]{babel}
\usepackage{amssymb}
\usepackage{amsmath}
\usepackage{mathtools}
\usepackage{txfonts}
\usepackage{mathdots}
\usepackage[normalem]{ulem}
\usepackage{array}
\usepackage{amssymb}
\usepackage{amsfonts}
\usepackage{amsmath}
\usepackage{mathrsfs}
\usepackage{booktabs}
\usepackage{threeparttable}
\usepackage{multirow}
\usepackage[dvips]{graphicx}
\usepackage{epsfig}
\usepackage{graphicx}
\usepackage{subfigure}
\usepackage{threeparttable}
\usepackage{chngpage}
\usepackage{float}
\usepackage[dvipsnames,svgnames,x11names,hyperref]{xcolor}
\usepackage{bm}
\usepackage[toc]{appendix}
\usepackage{tabularx}
\usepackage{adjustbox}
\usepackage{comment}
\usepackage{todonotes}
\usepackage{lineno}
\usepackage[unicode=true,bookmarks=true,bookmarksnumbered=false,bookmarksopen=false,breaklinks=false,pdfborder={0 0 1},backref=false,colorlinks=true]{hyperref}
\hypersetup{
 linkcolor=black, urlcolor=blue, citecolor=blue, pdfstartview={FitH}, unicode=true}
 \hypersetup{urlcolor=blue}

\newcommand{\dif}{\mathop{}\!\mathrm{d}}

\newcommand{\ave}[1]{\langle #1 \rangle}
\newcommand{\p}{\partial}
\newcommand{\T}{^{\mathrm{T}}}

\renewcommand{\vec}[1]{\boldsymbol{#1}}
\newcommand{\bnabla}{\vec{\nabla}}
\newcommand{\bn}{\bnabla}

\newcommand{\Od}[1]{\mathcal{O}( #1 )}
\newcommand{\dx}{\Delta x}
\newcommand{\dy}{\Delta y}

\renewcommand{\ge}{\geqslant}

\newcommand{\ket}[1]{| #1 \rangle}
\newcommand{\bra}[1]{\langle #1 |}

\newcommand{\beginsupplement}{%
	\setcounter{table}{0}
	\renewcommand{\thetable}{S\arabic{table}}%
	\setcounter{figure}{0}
	\renewcommand{\thefigure}{S\arabic{figure}}%
	\setcounter{equation}{0}
	\renewcommand{\theequation}{S\arabic{equation}}%
	\setcounter{section}{0}
	\renewcommand{\thesection}{\arabic{section}}%
}

\newtheorem{proof}{Proof}

\begin{document}
\title{
Simulating unsteady fluid flows on a superconducting quantum processor
	}

\author{Zhaoyuan Meng}\thanks{These authors contributed equally}
\affiliation{State Key Laboratory for Turbulence and Complex Systems, College of Engineering, Peking University, Beijing 100871, China}
\author{Jiarun Zhong}\thanks{These authors contributed equally}
\author{Shibo Xu}\thanks{These authors contributed equally}
\author{Ke Wang}
\author{Jiachen Chen}
\author{Feitong Jin}
\author{Xuhao Zhu}
\author{Yu Gao}
\author{Yaozu Wu}
\author{Chuanyu Zhang}
\author{Ning Wang}
\author{Yiren Zou}
\author{Aosai Zhang}
\author{Zhengyi Cui}
\author{Fanhao Shen}
\author{Zehang Bao}
\author{Zitian Zhu}
\author{Ziqi Tan}
\author{Tingting Li}
\author{Pengfei Zhang}
\affiliation{School of Physics, ZJU-Hangzhou Global Scientific and Technological Innovation Center, and Zhejiang Key Laboratory of Micro-nano Quantum Chips and Quantum Control, Zhejiang University, Hangzhou 310027, China}
\author{Shiying Xiong}
\affiliation{Department of Engineering Mechanics, School of Aeronautics and Astronautics, Zhejiang University, Hangzhou 310027, China}
\author{Hekang Li}
\author{Qiujiang Guo}
\author{Zhen Wang}
\author{Chao Song}
\email{chaosong@zju.edu.cn}
\author{H. Wang}
\affiliation{School of Physics, ZJU-Hangzhou Global Scientific and Technological Innovation Center, and Zhejiang Key Laboratory of Micro-nano Quantum Chips and Quantum Control, Zhejiang University, Hangzhou 310027, China}
\author{Yue Yang}
\email{yyg@pku.edu.cn}
\affiliation{State Key Laboratory for Turbulence and Complex Systems, College of Engineering, Peking University, Beijing 100871, China}
\affiliation{HEDPS-CAPT, Peking University, Beijing 100871, China}

\begin{abstract}
Recent advancements of intermediate-scale quantum processors have triggered tremendous interest in the exploration of practical quantum advantage.
The simulation of fluid dynamics, a highly challenging problem in classical physics but vital for practical applications, emerges as a good candidate for showing quantum utility.  
Here, we report an experiment on the digital simulation of unsteady flows, which consists of quantum encoding, evolution, and detection of flow states, with a superconducting quantum processor. 
The quantum algorithm is based on the Hamiltonian simulation using the hydrodynamic formulation of the Schr\"{o}dinger equation. 
With the median fidelities of $99.97\%$ and $99.67\%$ for parallel single- and two-qubit gates respectively, 
we simulate the dynamics of a two-dimensional (2D) compressible diverging flow and a 2D decaying vortex with ten qubits.
The experimental results well capture the temporal evolution of averaged density and momentum profiles, and qualitatively reproduce spatial flow fields with moderate noises.  
This work demonstrates the potential of quantum computing in simulating more complex flows, such as turbulence, for practical applications.
\end{abstract}

\maketitle 

\textit{Introduction.}---Simulating fluid dynamics on classical computers at a high Reynolds number (Re) has significant applications in various fields, such as weather forecasting and airplane design. However, it remains challenging due to the wide range of spatial and temporal scales involved in turbulent flows. 
Its computational cost, scaling with $\mathcal{O}(\mathrm{Re}^3)$ operations for the direct numerical simulation of turbulence~\cite{Pope2000_Turbulent}, is prohibitively expensive for engineering applications~\cite{Moin1998_Direct, Ishihara2009_Study}. 
The emergence of quantum computing has garnered attention as a potential solution to the computational limitations in classical computing~\cite{Manin1980_computable, Benioff1980_computer, Feynman1982_Simulating, Deutsch1985_quantum}. 
Leveraging laws of quantum mechanics such as superposition and entanglement, a quantum processor can manipulate exponentially large degrees of freedom that are intractable on classical computers, making it a promising platform for empowering the next-generation simulation method for fluid dynamics~\cite{Givi2020_Quantum, Succi2023_Quantum, Bharadwaj2023_Hybrid}.
In particular, quantum computing of turbulence, one of the most challenging problems in classical physics~\cite{Feynman2015_The}, can serve as a compelling demonstration of quantum utility and practical quantum advantage~\cite{Daley2022_Practical, Kim2023_Evidence, Hibat-Allah2024_A, Begusic2024_Fast}. 

There have been two major approaches to the quantum simulation of fluid dynamics. 
Based on solving the governing equations for fluids, hybrid quantum-classical algorithms are proposed~\cite{Steijl2018_Parallel, Gaitan2020_Finding, Chen2022_Quantum, Lapworth2022_A, Demirdjian2022_Variational, Gourianov2022_A, Pfeffer2022_Hybrid, Pfeffer2023_Reduced, Jaksch2023_Variational, Liu2023_Quantum, Bharadwaj2023_Hybrid, Succi2024_Ensemble, Au-Yeung2024_Quantum}, where quantum computing is employed to handle highly parallelizable operations (e.g., solving linear systems~\cite{Harrow2009_Quantum, Costa2022_Optimal}).
The efficiencies of these methods are often burdened by the frequent data exchanges between classical and quantum hardwares, as the preparation and statistical measurement of arbitrary quantum state can be more time-consuming than the calculation procedure~\cite{Succi2023_Quantum, Aaronson2015_Read}.
Moreover, for the present noisy intermediate-scale quantum (NISQ) devices, the state preparation and measurement (SPAM) errors could accumulate during the time-marching in these algorithms, limiting their accuracy for near-term applications~\cite{Preskill2018_Quantum, Bharti2022_Noisy}.
To alleviate these problems, Hamiltonian simulation, which has been widely used in exploring quantum many-body physics on NISQ devices~\cite{Bharti2022_Noisy, Georgescu2014_quantum, zhang_digital_2022, mi_time-crystalline_2022, Deng2022_Observing, Kim2023_Evidence}, was proposed as a promising approach to simulate fluid dynamics~\cite{Madelung1927_Quantentheorie, Yepez2001_Quantum, Joseph2020_Koopman, Budinski2021_Quantum, Zylberman2022_Quantum, Lu2023_Quantum, Meng2023_Quantum, Meng2023_Lagrangian, Meng2024_Quantum, Itani2024_Quantum}.
In this simulation, a fluid flow is mapped to a quantum system, which can then be evolved and detected on a quantum processor without invoking intermediate quantum state measurement and re-initialization.

However, obstacles remain. 
First, general fluid dynamics has nonlinear characteristics, while quantum operations except measurement are linear. Incorporating the nonlinearity into a quantum algorithm poses significant challenges.
Second, while minimizing the influence of SPAM errors, the Hamiltonian simulation can still be affected by the inevitable errors occurred during the execution of quantum evolution on the NISQ devices.
A proof-of-principle demonstration of the capability of the contemporary NISQ devices in simulating fluid dynamics remains elusive.
Here we report the quantum simulation of two-dimensional (2D) unsteady flows, discretized spatially with up to 1024 grid points, on a superconducting quantum processor.  
We first consider a simple compressible diverging flow and reveal its dynamics according to the hydrodynamic formulation of the corresponding Schr\"{o}dinger equation.
Then, by involving the two-component wave function, we realize the quantum simulation of a decaying vortex with nonlinear vortex dynamics.

\begin{figure}
    \centering
    \includegraphics[scale=1]{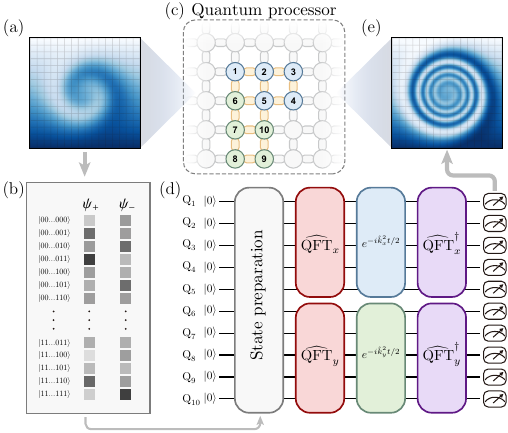}
    \caption{Schematic for the quantum simulation of fluid dynamics. The initial flow field (exemplified pictorially by a spiral vortex) discretized on a uniform grid in (a) is encoded by the multi-component wave function in (b), followed by state preparation.
    (c) Sketch of device topology. Qubits (circles) are arranged in a square lattice and connected through tunable couplers (bars). The ten qubits used here are labeled by Q$_1$--Q$_5$ (blue) and Q$_6$--Q$_{10}$ (green), encoding the wave function in the $x$- and $y$-directions, respectively. 
    (d) Sketch of the quantum circuit for simulating the 2D unsteady flows. The Hamiltonian simulation is realized by transforming a flow state into the momentum space with quantum Fourier transforms $\widehat{\mathrm{QFT}}_x$ and $\widehat{\mathrm{QFT}}_y$, applying unitary evolutions $e^{-i\hat{k}_x^2t/2}$ and $e^{-i\hat{k}_y^2t/2}$, and finally transforming the state back to the coordinate space.
    The circuit is further compiled with native gate sets (arbitrary single-qubit gates and two-qubit CZ gate) before the experimental execution.
    (e) At a given time $t$, the flow field is extracted by measuring a specific set of Pauli strings.
    }
    \label{fig:figure1}
\end{figure}

As sketched in Fig.~\ref{fig:figure1}, the simulation (referred to as ``experiment'' below) is implemented with ten qubits on a superconducting processor~\cite{Xu2023_digital, xu2024nonabelian, bao2024schrodinger}.
Through optimizing device fabrication and carefully tuning controlling parameters, we realize median fidelities of $99.97 \%$ ($99.67 \%$) and $99.3 \%$ for parallel single- (two-) qubit gates and measurements, respectively. With the high fidelities and by employing efficient quantum circuits for state preparation and Hamiltonian simulation, we obtain the average density and momentum profiles that well capture key features of the targeting flows. 
The present study demonstrates the capability of NISQ devices to simulate practical fluid flows, indicating the potential of quantum computing in exploring turbulent flows. 

\textit{Framework and experimental setup.}---In our algorithm, we encode the flow state into the $n_\psi$-component wave function $\vec{\psi} \equiv [\psi_1,\cdots,\psi_{n_\psi}]\T$, with $n_\psi\in\{1,2\}$. Based on the generalized Madelung transform, the flow density and momentum can be extracted as $\rho\equiv\sum_{j=1}^{n_\psi}|\psi_j|^2$ and $\vec{J}\equiv\frac{i\hbar}{2m}\sum_{j=1}^{n_\psi}(\psi_j\bn\psi_j^* - \psi_j^*\bn\psi_j)$, respectively~\cite{Madelung1927_Quantentheorie, Meng2023_Quantum, Meng2024_Quantum}. Without loss of generality, we set the reduced Planck constant $\hbar=1$ and the particle mass $m=1$. 
The fluid velocity and vorticity are defined by $\vec{u}\equiv \vec{J}/\rho$ and  $\vec{\omega}\equiv\bn\times\vec{u}$, respectively. 
Note that for a single-component wave function $\psi$, the velocity can be expressed as $\vec{u}=\frac{i}{2}\bn\log{\frac{\psi^*}{\psi}}$, leading to $\vec{\omega}=\vec{0}$.
To introduce finite vorticity, we need $n_\psi = 2$ \cite{Meng2023_Quantum}. 
We simulate fluid dynamics by evolving the wave function under the Hamiltonian $H = -\nabla^2/2 + V$, where the potential $V$, which may contain interaction terms among different wave-function components, gives the body force in the fluid flow~\cite{Meng2023_Quantum}. Using the Trotter decomposition~\cite{Trotter1959_Onthe}, the evolution of the wave function can be approximated by a series of unitary operators. 

\begin{figure*}
    \centering
    \includegraphics[scale=1]{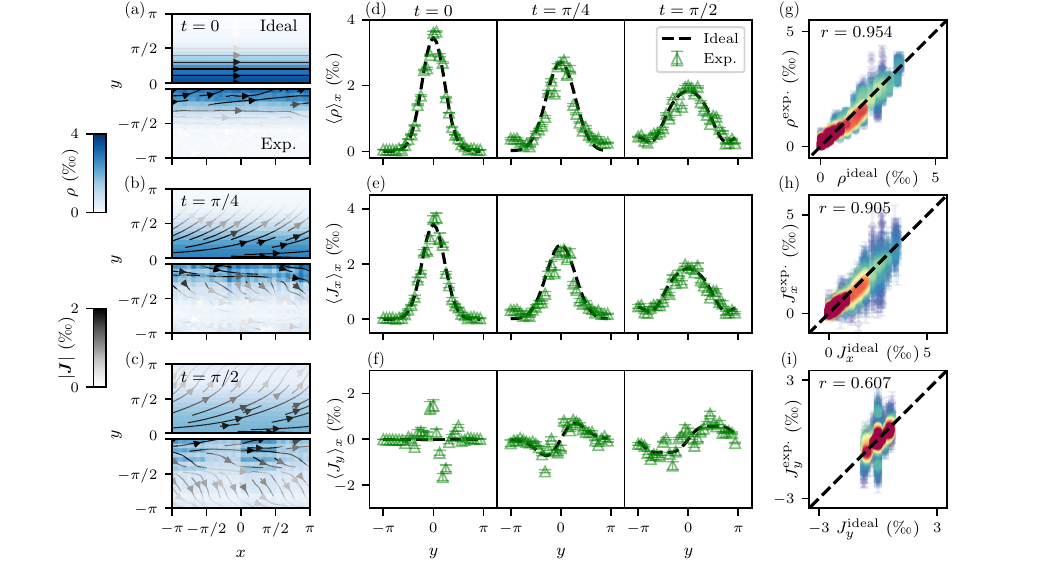}
    \caption{Comparison of experimental results on the superconducting quantum processor with the ideal ones for the 2D diverging flow. 
    Density contours and streamlines are shown at (a) $t=0$, (b) $t=\pi/4$, and (c) $t=\pi/2$. The flow is symmetric about $y=0$ (upper panel: exact solution; lower panel:  experimental measurement). The streamlines are color-coded by the momentum magnitude. 
    (d--f) The $x$-averaged profiles for $\rho$, $J_x$, and $J_y$ at $t=0$, $\pi/4$, and $\pi/2$ (dashed lines: exact solution; triangles: experimental measurement with error bars denoting one standard deviation). The density and momentum are obtained with $10^5$ measurement shots, and the experiment is repeated for five times. 
    (g--i) Scatter plots comparing ideal and experimental values for $\rho$, $J_x$, and $J_y$ at $t=0$, $\pi/4$, and $\pi/2$, along with the correlation coefficients (marked at the upper left). Data point density is color-coded from low (purple) to high (red) values.}
    \label{fig:figure2}
\end{figure*}

In this work, we focus on the dynamics of 2D flows without conservative body forces in a periodic box $\vec{x}\in[-\pi,\pi]^2$, which is discretized into $2^{n_x}\times2^{n_y}$ uniform grid points.
The corresponding wave function of each component can be expressed in the computational basis of $n_x+n_y$ qubits as
\begin{equation}
    \ket{\psi_j(t)} = \frac{1}{\|\psi_j\|_2}\sum_{l=0}^{2^{n_y}-1}\sum_{k=0}^{2^{n_x}-1}\psi_j(x_k,y_l,t)\ket{k+2^{n_x}l},
\end{equation}
where the coordinates $x_k=-\pi+k\dx$ and $y_l=-\pi+l\dy$, with $\dx=2\pi/2^{n_x}$ and $\dy=2\pi/2^{n_y}$, respectively. For vortical flows, an additional qubit is required to encode the two-component wave function. In the absence of conservative body forces, the Hamiltonian reduces to $H=-(\p_x^2+\p_y^2)/2$. The corresponding evolution can be realized without Trotterization as
\begin{equation}\label{eq:evolution}
    e^{-iHt} = e^{i\p_x^2t/2}e^{i\p_y^2t/2}=U_x(t)U_y(t).
\end{equation}
In the computational basis, the evolution operators can be digitized according to 
\begin{equation}\label{eq:fourier}
    U_\alpha(t) = \widehat{\mathrm{QFT}}_\alpha^\dagger e^{-i \hat{k}_\alpha^2t/2}\widehat{\mathrm{QFT}}_\alpha,
\end{equation}
where $\widehat{\mathrm{QFT}}_\alpha$ denotes the unitary of quantum Fourier transform along the direction $\alpha\in\{x, y\}$, and the wavenumber is given by
\begin{equation}
    \hat{k}_\alpha
    = -\frac{1}{2}\bigg(I_{2^{n_\alpha}} + \sum_{j=1}^{n_\alpha}2^{n_\alpha-j}\hat{Z}_j \bigg) + 2^{n_\alpha}\hat{Z}_1.
\end{equation}
Here, $\hat{Z}_j$ denotes the Pauli operator of the $j$-th qubit and $I_{2^{n_\alpha}}$ the $2^{n_\alpha}\times 2^{n_\alpha}$ identity matrix. See Supplementary Material (SM)~\cite{SM} for detailed derivations. 
The overall quantum circuit for simulating 2D unsteady flows is shown in Fig.~\ref{fig:figure1}(d). 

We implement the algorithm on a flip-chip superconducting quantum processor~\cite{Xu2023_digital} using ten frequency-tunable transmon qubits labeled as Q$_j$ for $j=1$--$10$, as sketched in Fig.~\ref{fig:figure1}(c). In the simulation, we set $n_x=n_y=5$, corresponding to a solution domain with $2^5 \times 2^5 = 32^2$ grid points. Each qubit is individually controlled and readout. The nearest-neighboring qubits are capacitively connected through a tunable coupler, which is also a transmon qubit, for turning on and off the effective coupling between the two qubits. 
The single-qubit gate is realized by applying a 30 ns-long Gaussian-shaped microwave pulse with DRAG correction~\cite{PhysRevLett.112.240504}. 
The two-qubit CZ gate, with a length of 40 ns, is realized by carefully tuning the frequencies and coupling strength of the neighboring qubits to enable a closed-cycle diabatic transition of $|11\rangle\leftrightarrow|20\rangle$ (or $|02\rangle $)~\cite{Ren2022Experimental}.  
The median parallel single- and two-qubit gate fidelities are $99.97\%$ and $99.67\%$, respectively.
See SM~\cite{SM} for details.

\textit{Simulation of a diverging flow.}---As a first example, we demonstrate the quantum simulation of a 2D unsteady diverging flow, which is a simple model of nozzle in compressible potential flow.
The flow is initially uniform in the $x$-direction, with mass concentrated near $y=0$, which is described by a density of $\rho(x, y,0)=e^{-y^2}$ and a velocity of $\vec{u}(x, y, 0)=\vec{e}_x$.
The flow has the vanishing vorticity and can be encoded into a single-component wave function $\psi(x, y,0)=e^{-y^2+ix}$.

In practice, we use CPFlow~\cite{Nemkov2023_Efficient} to synthesize the quantum circuit for initial state preparation, which fits the native gate set (i.e., arbitrary single-qubit gates and two-qubit CZ gates) and qubit layout topology of our device with minimal numbers of CZ gates.
With an optimal depth of 13, the resulting circuit can generate a quantum state with an overlap above 0.999 to the target one in the ideal case (i.e., without any gate errors).
After preparing the initial state, we apply quantum circuits of the evolution unitaries with specific times of $t=0, \pi/4$ and $\pi/2$, respectively.
The evolution circuits are also optimized with CPFlow, leading to a total circuit depth of around {30}.
To verify the simulation, we fully characterize the flow by measuring both the density and momentum distributions.
While the density can be directly measured in the $Z$ basis, the detection of the momentum requires measuring the quantum state in 62 different bases (see SM~\cite{SM} for details).

Figures~\ref{fig:figure2}(a--c) show the experimental data for the evolution of the density contour with streamlines for the diverging flow. 
Under the symmetry with respect to $y=0$, the experimental data (lower half) is compared with the ideal result (upper half) from the exact solution in Eq.~\eqref{eq:SM_psi_exact_1} in SM~\cite{SM}.  
The mass diffusion accompanies the momentum diffusion from the central region near $y=0$ to lateral sides. 
The experimental data exhibit qualitative agreements with ideal distributions. 
The discrepancies are mainly due to quantum gate errors, as even an single-qubit gate error of $5\times10^{-4}$ can cause the stripe-like artifacts in experimental density contours (see SM~\cite{SM}).

Figures~\ref{fig:figure2}(d--f) plot profiles of $\rho$ as well as two momentum components $J_x$ and $J_y$. 
The profiles are averaged in the $x$-direction due to the homogeneity in $x$ in this diverging flow. 
Their experimental and ideal results show good agreements.  
Thus, the quantum simulation on the NISQ device is able to predict unsteady flow evolution. 

Figures~\ref{fig:figure2}(g--i) present scatter plots comparing ideal and experimental values of $\rho$, $J_x$, and $J_y$ at different times. 
A majority of data points, indicated by red for high data point density, align closely with the diagonal. 
In Figs.~\ref{fig:figure2}(g) and (h), a considerable portion of experimental data falls below the actual values (orange), 
suggesting the effect of noises akin to filtering on the data. 
The correlation coefficients between the experimental and ideal values of $\rho$, $J_x$, and $J_y$ are $0.954$, $0.905$, and $0.607$, respectively. 
The notable error for $J_y$ data is likely due to that the small value of $J_y$ is prone to be influenced by noises (detailed in SM~\cite{SM}). 

\textit{Simulation of a decaying vortex.}---Next, we simulate a 2D vortex, a simple model of tornado and whirlpool, in the Schrödinger flow~\cite{Meng2023_Quantum} (detailed in SM~\cite{SM}).
We construct the vortex via rational maps~\cite{Kedia2016_Weaving} in the periodic domain, which features a decaying vorticity profile $f(r)=e^{-(r/3)^4}$ along the radial distance $r=\sqrt{x^2+y^2}$.
Denoting the two components of the wave function as $\psi_+$ and $\psi_-$, the initial states $\psi_+(x, y,0)=u/\sqrt{|u|^2+|v|^4}$ and $\psi_-(x, y,0)=v^2/\sqrt{|u|^2+|v|^4}$ are determined by the complex functions $u=2(x+iy)f(r)/(1+r^2)$ and $v=i[r^2+1-f(r)]/(1+r^2)$, respectively. 

In this flow with $V=0$ in the Hamiltonian, the two components of the wave function decouple during the entire evolution, so that we can simulate their dynamics separately without using an additional qubit. 
The circuits for preparing $\psi_+$ and $\psi_-$, obtained based on CPFlow, have depths of 23 and 27, respectively, which can prepare quantum states with overlaps above 0.993 to the target ones in the ideal case.
We then evolve $\psi_+$ and $\psi_-$ independently for specific times and measuring the corresponding densities $\rho_\pm$ and momentum $\vec{J}_\pm$, with the procedure similar to the diverging flow case.
The velocity is then obtained by $\vec{u} = (\vec{J}_+ + \vec{J}_-)/(\rho_+ + \rho_-)$, which introduces the nonlinearity for vortex dynamics, and the vorticity is calculated as $\omega=\p u_y/\p x - \p u_x/\p y$. 

Figures~\ref{fig:figure3}(a) and (b) plot the evolution of the $\omega$-contour and streamlines from experimental and ideal results at $t=0$, $\pi/4$, and $\pi/2$. 
Initial circular streamlines evolve into spirals with vorticity decay under the body force in the Schrödinger flow. 
The experimental results in Fig.~\ref{fig:figure3}(b) clearly capture the vortex evolution. 
The vorticity magnitude is underestimated due to quantum gate noises and sampling errors. 
At the vortex outer edge, the flow field exhibits turbulent artifacts which enhances vortex dissipation.  
On the other hand, the NISQ hardware noises could potentially be leveraged to model small-scale turbulent motion~\cite{Pope2011_Simple}. 

\begin{figure}
    \centering
    \includegraphics[scale=1]{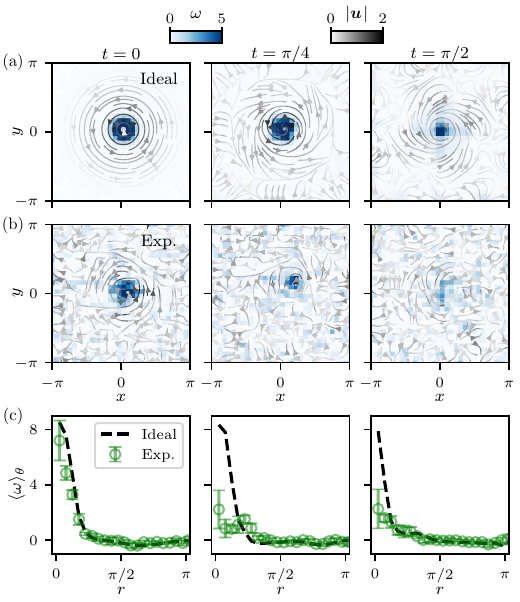}
    \caption{Comparison of vorticity contours from the (a) exact solution (b) experimental result for the 2D vortex. 
    Streamlines are color-coded by velocity magnitude. 
    (c) Comparison of $\langle\omega\rangle_\theta$ at $t=0$, $\pi/4$, and $\pi/2$ (dashed lines: exact solution; circles: experimental result with error bars denoting one standard deviation), where $\langle\omega\rangle_\theta$ denotes the vorticity averaged over the $\theta$-direction in polar coordinates $(r,\theta)$. The data is obtained with $10^5$ measurement shots, and the experiment is repeated for five times.}
    \label{fig:figure3}
\end{figure}

The $\theta$-averaged profiles for $\omega$ in Fig.~\ref{fig:figure3}(c) show the successful initial construction of the vortex using the two-component wave function. 
The peak of $\ave{\omega}_\theta$ decays notably faster in the experimental results compared to the exact solution, due to the vortex being displaced from the domain center under spurious turbulent motion and the numerical error in computing $\omega$ with differentiation of noisy data.

\textit{Discussion.}---We have conducted experiments on the digital quantum simulation of unsteady fluid flows with a superconducting quantum processor.
Our algorithm employs the hydrodynamic formulation of the Schrödinger equation, apt for unitary operations in quantum computing. 
The computational complexity for the state evolution (detailed in SM~\cite{SM}) in Eq.~\eqref{eq:evolution} is $\mathcal{O}(n^2)$~\cite{Nielsen2010_Quantum}, with the total number of qubits $n=n_x + n_y$. 
This represents an exponential speedup over the classical counterpart whose complexity is $\mathcal{O}(n2^{n})$.  
The quantum simulations well capture the evolution of averaged profiles of the density and momentum in the 2D compressible diverging potential flow and the nonlinear decaying process of the 2D vortex. 
Our results showcase the capability of simulating fluid dynamics on NISQ devices, and indicate the promise of quantum computing in probing complex flow phenomena such as turbulence and transition in engineering applications. 

Looking forward, despite the demonstration in the present work, realizing quantum advantage for the simulation of practical fluid flows with NISQ devices remains an outstanding challenge. 
Theoretically, methods such as increasing dimensionality~\cite{Lloyd2020_Quantum, Jin2023_Quantum, Koukoutsis2023_Dyson} should be explored to incorporate the nonlinearity and the non-Hermitian Hamiltonian of a general flow~\cite{Ashida2021_Non, Meng2024_Quantum} into the quantum algorithm.
In addition, the full characterization of a flow field requires an exponentially large number of measurement shots, and it would be important to find flow statistics that can be measured efficiently without undermining the overall quantum advantage. 
Finally, the potential inclusion of quantum error correction is desired to fully harness the strengths of the quantum simulation of fluid dynamics. 

\smallskip
This work has been supported by the National Key R\&D Program of China (Grant No.~2023YFB4502600), the National Natural Science Foundation of China (Grant Nos.~11925201, 11988102, 12174342, 12274367, 12274368, 12322414, and 92365301), the Zhejiang Provincial Natural Science Foundation of China (Grant Nos.~LR24A040002 and LDQ23A040001), and the New Cornerstone Science Foundation through the XPLORER Prize. 

\let\oldaddcontentsline\addcontentsline
\renewcommand{\addcontentsline}[3]{}
%

\let\addcontentsline\oldaddcontentsline

\resetlinenumber
\clearpage
\onecolumngrid
\begin{center}
\textbf{\large Supplementary Material for \\
    ``Simulating unsteady fluid flows on a superconducting quantum processor''}
\end{center}

\maketitle
\setcounter{page}{1}

\beginsupplement
\renewcommand{\thepage}{S\arabic{page}}
\renewcommand{\citenumfont}[1]{S#1}
\renewcommand{\bibnumfmt}[1]{[S#1]}

\tableofcontents

\section{Quantum representation of fluid flows}
\subsection{Madelung transform for potential flows}
In quantum mechanics, the probability current for a one-component wave function $\psi(\vec{x},t)$ is given by
\begin{equation}\label{eq:SM_J}
    \vec{J}(\vec{x},t) \equiv \frac{1}{2m}(\psi^*\hat{\vec{p}}\psi - \psi\hat{\vec{p}}\psi^*),
\end{equation}
where $\hat{\vec{p}}$ is the momentum operator, $m$ is the particle mass, and ``$*$'' denotes the complex conjugate. 
In the coordinate representation, we have $\hat{\vec{p}}=-i\hbar\bn$ with the imaginary unit $i$ and Planck constant $\hbar$. 
A particle subject to a real potential $V$ obeys the time-dependent Schrödinger equation
\begin{equation}\label{eq:SM_Schrodinger}
    i\hbar\frac{\p}{\p t}\psi(\vec{x},t) = \bigg(-\frac{\hbar^2}{2m}\nabla^2 + V \bigg)\psi(\vec{x},t).
\end{equation}
Combining Eqs.~\eqref{eq:SM_J} and \eqref{eq:SM_Schrodinger} yields the conservation law for the probability density $\rho\equiv |\psi|^2$ as
\begin{equation}\label{eq:SM_continuity}
    \frac{\p\rho}{\p t} + \bn\cdot(\rho\vec{u}) = 0,
\end{equation}
which is identical to the continuity equation in fluid mechanics, with a ``velocity'' $\vec{u}\equiv \vec{J}/\rho$.

The Madelung transform~\cite{SM_Madelung1927_Quantentheorie} establishes a correspondence between quantum mechanics and fluid mechanics.
The probability density and the probability current in quantum mechanics are analogous to the mass density and momentum in fluid mechanics, respectively. 
The Planck constant in this hydrodynamic representation is a scaling parameter instead of a physical constant.  

Applying the Madelung transform to Eq.~\eqref{eq:SM_Schrodinger} yields the momentum equation
\begin{equation}\label{eq:SM_momentum_classical}
    \frac{\p\vec{u}}{\p t} + \vec{u}\cdot\bn\vec{u} = -\frac{1}{m}\bn V + \frac{\hbar^2}{2m^2}\bn\frac{\nabla^2\sqrt{\rho}}{\sqrt{\rho}}
\end{equation}
for a fluid flow. 
The density and velocity are given by
\begin{equation}\label{eq:SM_rho_vel}
    \rho = |\psi|^2, \quad
    \vec{u} = \frac{i\hbar}{2m}\frac{\psi\bn\psi^* - \psi^*\bn\psi}{\rho}.
\end{equation}
The second term on the right-hand side of Eq.~\eqref{eq:SM_momentum_classical} is known as the Bohm potential. 
Substituting $\psi=\sqrt{\rho}e^{i\phi/\hbar}$ into Eq.~\eqref{eq:SM_rho_vel}, the velocity becomes $\vec{u}=\bn\phi/m$, so Eq.~\eqref{eq:SM_momentum_classical} is the compressible Euler equation for a potential flow with zero vorticity. 

\subsection{Generalized Madelung transform for vortical flows}
To introduce finite vorticity to a fluid flow, we consider a two-component wave function
\begin{equation}
    \ket{\psi_\pm} = \begin{bmatrix}
        \psi_+ \\ \psi_-
    \end{bmatrix},
\end{equation}
where $\psi_+$ and $\psi_-$ denote the complex spin-up and -down components, respectively~\cite{SM_Mueller2002_Two, SM_Kasamatsu2003_Vortex, SM_Meng2023_Quantum}.
The probabilistic current in Eq.~\eqref{eq:SM_J} is generalized to
\begin{equation}
    \vec{J} \equiv \frac{\hbar}{m}\langle \bn\psi_{\pm},i\psi_{\pm}\rangle_{\mathbb{R}},
\end{equation}
where $\langle\cdot,\cdot\rangle_{\mathbb{R}}$ is the real inner product defined as $\langle \phi, \psi\rangle_{\mathbb{R}}\equiv \sum_{i=1}^2 (x_i\alpha_i + y_i\beta_i)$ for $\phi=[x_1+i y_1, x_2+i y_2]\T$ and $\psi=[\alpha_1+i \beta_1, \alpha_2+i\beta_2]\T$.
Using the generalized Madelung transform~\cite{SM_Meng2023_Quantum}, we define the flow velocity
\begin{equation}\label{eq:SM_vel}
    \vec{u} \equiv \frac{\vec{J}}{\rho}
    = \frac{\hbar}{m}\frac{\langle \bn\psi_{\pm},i\psi_{\pm}\rangle_{\mathbb{R}}}{\langle\psi_{\pm}, \psi_{\pm}\rangle_{\mathbb{R}}},
\end{equation}
where
\begin{equation}\label{eq:SM_rho}
    \rho
    \equiv \langle\psi_{\pm}, \psi_{\pm}\rangle_{\mathbb{R}}
    = |\psi_+|^2+|\psi_-|^2
\end{equation} 
is the mass density.
The flow field given by Eq.~\eqref{eq:SM_vel} has finite vorticity, i.e., $\vec{\omega}\equiv\bn\times\vec{u}\ne\vec{0}$.

Then, we consider a two-component time-dependent Schr\"odinger equation
\begin{equation}\label{eq:SM_two_Schr}
    i\hbar\frac{\p}{\p t}\ket{\psi_\pm} = \bigg(-\frac{\hbar^2}{2m}\nabla^2 + V \bigg)\ket{\psi_\pm}
\end{equation}
with a real-valued potential $V$.
The continuity equation follows from the conservation of probability current. 
The momentum equation is obtained as~\cite{SM_Meng2023_Quantum} 
\begin{equation}\label{eq:SM_pupt}
    \frac{\p\vec{u}}{\p t} + \vec{u}\cdot\bn\vec{u}
    = -\bn \bigg(\frac{V}{m} - \frac{\hbar^2|\bn\vec{s}|^2}{8m^2\rho^2} \bigg) -\frac{\hbar}{m\rho}\bn(\vec{\zeta}\cdot\vec{s}) + \frac{\hbar}{m\rho}\bn\vec{s}\cdot\vec{\zeta},
\end{equation}
where we define the spin vector
\begin{equation}
    \vec{s} = (|\psi_+|^2-|\psi_-|^2, i(\psi_+^*\psi_- - \psi_+\psi_-^*), \psi_+^*\psi_- + \psi_+\psi_-^*),
\end{equation}
and a vector
\begin{equation}
    \vec{\zeta} = -\frac{1}{4}\bn\cdot\bigg(\frac{\bn\vec{s}}{\rho} \bigg).
\end{equation}

Equation~\eqref{eq:SM_pupt} can be re-expressed as a standard compressible Euler equation 
\begin{equation}\label{eq:SM_vor_Euler}
    \frac{\p\vec{u}}{\p t} + \vec{u}\cdot\bn\vec{u}
    = -\frac{1}{\rho}\bn p - \bn U_F + \vec{f},
\end{equation}
with the ``quantum pressure'' $p=\hbar\vec{\zeta}\cdot\vec{s}/m$, an effective potential $U_F=V/m-\hbar^2|\bn\vec{s}|^2/(8m^2\rho^2)$, and a body force $\vec{f}=\hbar\bn\vec{s}\cdot\vec{\zeta}/(m\rho)$.

\section{Construction of quantum circuits}
\subsection{Initial state preparation}
We use two typical simple flows to demonstrate the capability of quantum simulation of fluid dynamics on the NISQ device. 
Their initial condition and evolution are described below. 

The first case is a 2D diverging compressible potential flow within a periodic domain $\vec{x} \in [-\pi, \pi]^2$. 
The initial wave function is
\begin{equation}\label{eq:SM_initial_condition_1}
    \psi(\vec{x},0)=e^{-y^2/(2\varrho^2)+i x},
\end{equation}
with a geometric parameter $\varrho=1$. 
For clarity, we omit the normalization constant in Eq.~\eqref{eq:SM_initial_condition_1} for the unity encoded quantum state. 
Applying the Madelung transform to Eq.~\eqref{eq:SM_initial_condition_1} yields the initial density and velocity
\begin{equation}
    \rho(\vec{x},0) = e^{-y^2/\varrho^2} \quad
    \text{and} \quad
    \vec{u}(\vec{x},0) = \vec{e}_x,
\end{equation}
where $\{\vec{e}_x, \vec{e}_y\}$ are the Cartesian unit vectors. 
The initial condition in Eq.~\eqref{eq:SM_initial_condition_1} describes a flow which is uniform in the $x$-direction and has a Gaussian profile of the density in the $y$-direction. 
The thickness of the high-density layer around $y=0$ is determined by $\varrho$. 

Applying the Fourier transform, the time integration for the initial value problem in spectral space, and then the inverse Fourier transform, we obtain the exact solution to Eq.~\eqref{eq:SM_Schrodinger} as
\begin{align}
    \psi(\vec{x},t) &= \frac{1}{(2\pi)^2}\sum_{\vec{k}}\bigg(\int_{[-\pi,\pi]^2}e^{-y^2/(2\varrho^2)+i x}e^{-i\vec{k}\cdot\vec{x}}\dif\vec{x}\bigg)e^{-i k^2 t/2 + i\vec{k}\cdot\vec{x}}
    \notag \\
    &= \frac{\varrho}{\sqrt{8\pi}}\sum_{k_2=-\infty}^\infty\bigg[\mathrm{erf}\bigg(\frac{\pi - i\varrho k_2}{\sqrt{2}\varrho}\bigg) + \mathrm{erf}\bigg(\frac{\pi + i\varrho k_2}{\sqrt{2}\varrho}\bigg) \bigg] e^{-\varrho^2k_2^2/2 -i(k_2^2+1)t/2 + i(x+k_2y)},
    \label{eq:SM_psi_exact_1}
\end{align}
where $\vec{k}=(k_1,k_2),~k_i=0,\pm1,\pm2,\cdots,~i=1,2$ denotes the wavenumber vector, $k=|\vec{k}|$ the wavenumber magnitude, and $\mathrm{erf}(z)\equiv \frac{2}{\sqrt{\pi}}\int_0^ze^{-t^2}\dif t$ the error function. 

The second case is a vortical flow with an artificial body force governed by Eq.~\eqref{eq:SM_vor_Euler}. 
We apply the rational maps~\cite{SM_Kedia2016_Weaving} to construct a 2D vortex at the center of the periodic domain $\vec{x}\in[-\pi,\pi]^2$. 
First, we define the vortex decay rate by
\begin{equation}
    f(r) = e^{-(r/r_0)^4},
\end{equation}
with the radial distance $r=\sqrt{x^2+y^2}$ and the vortex size $r_0=3$. 
Then, we define a pair of complex-valued functions
\begin{equation}
    u = \frac{2(x + i y)f(r)}{1+r^2}, \quad
    v = \frac{i[r^2 + 1 - 2f(r)]}{1 + r^2}.
\end{equation}
Finally, we specify the initial spin-up and -down components
\begin{equation}\label{eq:SM_initial_condition_2}
    \psi_+(\vec{x},0) = \frac{u}{\sqrt{|u|^2+|v|^4}}, \quad \psi_-(\vec{x},0) = \frac{v^2}{\sqrt{|u|^2+|v|^4}}.
\end{equation}
Note that we omit the normalization constant in Eq.~\eqref{eq:SM_initial_condition_2}.

The exact solutions to Eq.~\eqref{eq:SM_two_Schr} are
\begin{equation}\label{eq:SM_exact_psi1}
    \psi_+(\vec{x},t)
    = \frac{1}{(2\pi)^2}\sum_{\vec{k}}\bigg(\int_{[-\pi,\pi]^2}\frac{u}{\sqrt{|u|^2+|v|^4}}e^{-i\vec{k}\cdot\vec{x}}\dif\vec{x} \bigg)e^{-i k^2t/2 + i\vec{k}\cdot\vec{x}},
\end{equation}
and
\begin{equation}\label{eq:SM_exact_psi2}
    \psi_-(\vec{x},t)
    = -\frac{1}{(2\pi)^2}\sum_{\vec{k}}\bigg(\int_{[-\pi,\pi]^2}\frac{v^2}{\sqrt{|u|^2+|v|^4}}e^{-i\vec{k}\cdot\vec{x}}\dif\vec{x} \bigg)e^{-i k^2t/2 + i\vec{k}\cdot\vec{x}}.
\end{equation}
With Eqs.~\eqref{eq:SM_exact_psi1} and \eqref{eq:SM_exact_psi2}, we obtain the velocity field from Eq.~\eqref{eq:SM_vel}. 
Subsequently, we calculate the vorticity by computing the curl of this velocity field. 

\begin{figure}
    \centering
    \includegraphics[width=0.7\linewidth]{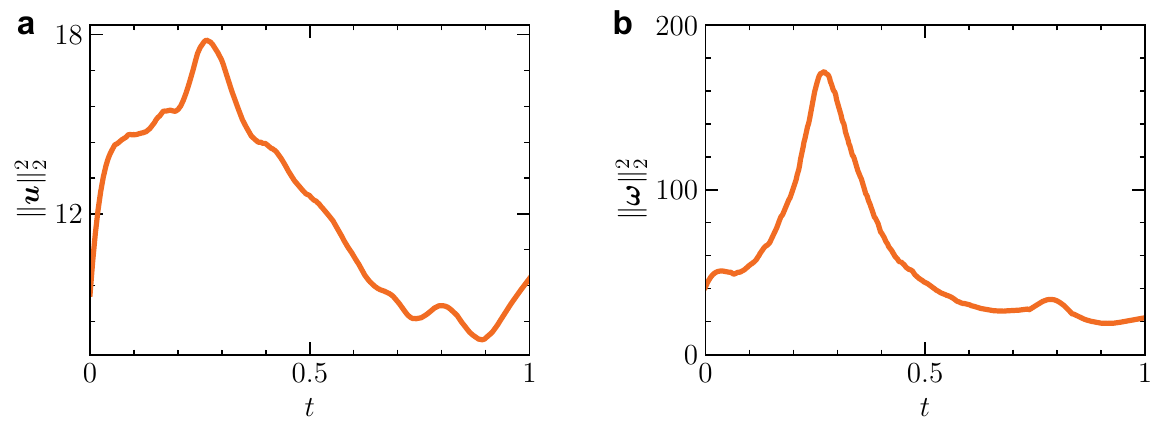}
    \caption{Evolution of (a) the total kinetic energy and (b) enstrophy for the 2D vortex in Eqs.~\eqref{eq:SM_exact_psi1} and \eqref{eq:SM_exact_psi2}.}
    \label{fig:vortex_statistics}
\end{figure}

Figure~\ref{fig:vortex_statistics} shows the evolution of the total kinetic energy and enstrophy for the 2D vortex in Eqs.~\eqref{eq:SM_exact_psi1} and \eqref{eq:SM_exact_psi2}. 
Due to the artificial body force, they both initially increase and subsequently decreases. 

\subsection{Evolution of quantum state}
For potential flows, the wave function is encoded by the state vector
\begin{equation}
    \ket{\psi(t)} = \frac{1}{\|\psi\|_2}\sum_{k=0}^{N_y-1}\sum_{j=0}^{N_x-1}\psi(x_j,y_k,t)\ket{j,k},
\end{equation}
where $N_x=2^{n_x}$ and $N_y=2^{n_y}$ denote numbers of grid points along the $x$- and $y$-directions, corresponding to numbers $n_x$ and $n_y$ of qubits, respectively, and
\begin{equation}
    \|\psi\|_2 = \sqrt{\sum_{j=0}^{N_x-1}\sum_{k=0}^{N_y-1}|\psi(x_j,y_k,t)|^2}
\end{equation}
represents the normalization coefficient. 
To avoid deep circuits due to iterations in time marching, we impose $V=0$ in Eq.~\eqref{eq:SM_Schrodinger}, signifying the absence of conservative body forces in the fluid flow.
This condition allows the time step to be arbitrarily large, enabling a ``one-shot'' solution \cite{SM_Lu2023_Quantum} for a given time $t$ without temporal discretization and intermediate measurement.  

Thus, the state vector at $t$ is determined by
\begin{equation}
    \ket{\psi(t)}
    = \left(\widehat{\mathrm{QFT}}_x^\dagger e^{-i \hat{k}_x^2t/2}\widehat{\mathrm{QFT}}_x\right)  \otimes \left(\widehat{\mathrm{QFT}}_y^\dagger e^{-i \hat{k}_y^2t/2}\widehat{\mathrm{QFT}}_y\right)\ket{\psi(0)},
\end{equation}
with the quantum Fourier transform (QFT)~\cite{SM_Jozsa1998_Quantum, SM_Weinstein2001_Implementation}
\begin{equation}
    \widehat{\mathrm{QFT}}:~ \ket{j}\to\frac{1}{\sqrt{2^n}}\sum_{k=0}^{2^n-1}e^{2\pi i\frac{jk}{2^n}}\ket{k},
\end{equation}
the wavenumber vector $\hat{\vec{k}}=(\hat{k}_x,\hat{k}_y)$, and
\begin{equation}\label{eq:SM_k_0}
    \hat{k}_\alpha=\mathrm{diag}(0,1,2,\cdots,2^{n_\alpha-1}-2,2^{n_\alpha-1}-1,-2^{n_\alpha-1},-2^{n_\alpha-1}+1,\cdots,-1),~\alpha=x,y.
\end{equation}
We rewrite Eq.~\eqref{eq:SM_k_0} as
\begin{equation}\label{eq:SM_k_1}
    \hat{k}_\alpha = \mathrm{diag}(0,1,\cdots,2^{n_\alpha}-1) - 2^{n_\alpha}\begin{bmatrix}
        0 & 0 \\ 0 & 1
    \end{bmatrix}\otimes I_{2^{n_\alpha-1}}
    = \mathrm{diag}(0,1,\cdots,2^{n_\alpha}-1) - 2^{n_\alpha-1}\left(I_{2^{n_\alpha}} - \hat{Z}\otimes I_{2^{n_\alpha-1}} \right),
\end{equation}
where $\hat{Z}=\mathrm{diag}(1,-1)$ denotes the Pauli-$Z$ gate, and $I_{2^{n_\alpha}}$ the $2^{n_\alpha}\times 2^{n_\alpha}$ identity matrix.

Then, we prove the \textit{proposition}
\begin{equation}\label{eq:SM_proposition}
    P(n_\alpha):~\mathrm{diag}(0,1,\cdots,2^{n_\alpha}-1) = \frac{1}{2}\bigg[(2^{n_\alpha}-1)I_{2^{n_\alpha}} - \sum_{j=1}^{n_\alpha}2^{n_\alpha-j}I_{2^{j-1}}\otimes\hat{Z}\otimes I_{2^{n_\alpha-j}} \bigg]
\end{equation}
by mathematical induction. 
\begin{proof}
    Base case: for $n_\alpha=1$, the identity
    \begin{equation}
        \mathrm{diag}(0,1) = \frac{1}{2}(I_2 - \hat{Z})
    \end{equation}
    verifies $P(1)$.
    
    Induction step: we will show $P(k+1)$ from $P(k) $ for any positive integer $k$. 
    Assume for a given value $k\ge 1$, the single case $P(k)$ is true, i.e., 
    \begin{equation}
        \mathrm{diag}(0,1,\cdots,2^{k}-1) = \frac{1}{2}\bigg[(2^{k}-1)I_{2^{k}} - \sum_{j=1}^{k}2^{k-j}I_{2^{j-1}}\otimes\hat{Z}\otimes I_{2^{k-j}} \bigg].
    \end{equation}
    For $k+1$, we deduce
    \begin{align}
        \mathrm{diag}(0,1,\cdots,2^{k+1}-1)
        &= \begin{bmatrix}
            \mathrm{diag}(0,1,\cdots,2^{k}-1) & \vec{0} \\ 
            \vec{0} & \mathrm{diag}(0,1,\cdots,2^{k}-1) + 2^{k}I_{2^k}
        \end{bmatrix}
        \notag \\
        &= I_2\otimes\mathrm{diag}(0,1,\cdots,2^{k}-1) + 2^k\begin{bmatrix}
            0 & 0 \\ 0 & 1
        \end{bmatrix}\otimes I_{2^k}
        \notag \\
        &= \frac{1}{2}\bigg[(2^k-1)I_{2^{k+1}} - \sum_{j=2}^{k+1}2^{k+1-j}I_{2^{j-1}}\otimes\hat{Z}\otimes I_{2^{k+1-j}} \bigg] + 2^k\frac{I_2-\hat{Z}}{2}\otimes I_{2^k}
        \notag \\
        &= \frac{1}{2}\bigg[(2^{k+1}-1)I_{2^{k+1}} - \sum_{j=1}^{k+1}2^{k+1-j}I_{2^{j-1}}\otimes\hat{Z}\otimes I_{2^{k+1-j}} \bigg].
    \end{align}
    So $P(k+1)$ is true, and the proposition $P(n_\alpha)$ in Eq.~\eqref{eq:SM_proposition} is true for all positive integers $n_\alpha$.
\end{proof}

Substituting Eq.~\eqref{eq:SM_proposition} into Eq.~\eqref{eq:SM_k_1}, we obtain
\begin{equation}\label{eq:SM_k}
    \hat{k}_\alpha
    = -\frac{1}{2}\bigg(I_{2^{n_\alpha}} + \sum_{j=1}^{n_\alpha}2^{n_\alpha-j}\hat{Z}_j \bigg) + 2^{n_\alpha}\hat{Z}_1,~\alpha=x,y,
\end{equation}
with the phase-shift gate $\hat{Z}_j=I_{2^{j-1}}\otimes\hat{Z}\otimes I_{2^{n_\alpha-j}}$ at the $j$-th qubit.

Then, from Eq.~\eqref{eq:SM_k}, the momentum operator
\begin{equation}
    e^{-i\hat{k}_\alpha^2t/2}
    = \exp\bigg\{-\frac{it}{2}\bigg[-\frac{1}{2}\bigg(I_{2^{n_\alpha}} + \sum_{j=1}^{n_\alpha}2^{n_\alpha-j}\hat{Z}_j \bigg) + 2^{n_\alpha}\hat{Z}_1 \bigg]^2 \bigg\}
\end{equation}
is decomposed into the single-qubit gate
\begin{equation}
    R_z(\theta) \equiv e^{-i\theta\hat{Z}/2}
    = \begin{bmatrix}
        e^{-i\theta/2} & 0 \\
        0 & e^{i\theta/2}
    \end{bmatrix}
 \end{equation}
and the two-qubit gate
\begin{equation}
    \mathrm{ZZ}(\theta) \equiv e^{-i\theta \hat{Z}\otimes\hat{Z}/2}
    = \mathrm{diag}(e^{-i\theta/2},e^{i\theta/2},e^{i\theta/2},e^{-i\theta/2})
    = \mathrm{CNOT}[I_2\otimes R_z(\theta)]\mathrm{CNOT},
\end{equation}
where $\mathrm{CNOT}$ denotes the controlled-NOT gate.
Note that the global phase from $I_{2^{n_\alpha}}$ can be ignored.

For vortical flows, an additional qubit can be added to encode the spin state~\cite{SM_Meng2023_Quantum} into the $n_x+n_y+1$ qubits. 
Instead, we split the two components in $\psi_\pm$, and treat $\psi_+$ or $\psi_-$ as $\psi$ for potential flows, due to the uncoupled nature of the two components in Eq.~\eqref{eq:SM_two_Schr}.
This permits utilization of the identical qubit array as in potential flow and a shorter circuit for improving experimental outcomes. 

\subsection{Circuit optimization}
The original circuits for initial state preparation and evolution may contain CZ gates between arbitrary two qubits.
Constrained by the layout topology of our device, only CZ gates between adjacent qubits are allowed, necessitating further circuit transpilations.

In this work, considering the limited fidelities of the quantum gates on our device, we use CPFlow~\cite{SM_Nemkov2023_Efficient} to obtain approximate circuits that fit the native gate set and device topology, meanwhile minimize the number of CZ gates. 
This program takes connections between qubits, the threshold, and the loss function based on the targeting state or unitary matrix as main input. 
Here we choose $1-\mathrm{F}$ as the loss function where F is the fidelity between the target and result. 
Based on the input connections, it randomly generate entangling blocks containing the gate set $\{\mathrm{CP}, R_x, R_y, R_z\}$, where $\mathrm{CP}$ denotes controlled-phase gate and $R_i$ represents rotation gates around the axis $i$, and adjust angles of each gate to minimize the loss function. 
The optimization procedure ends when the threshold is reached. 
The fidelity of results generated by CPFlow is listed in Tab.~\ref{table1}. 

\begin{table}[h]
    \centering
    \caption{Fidelity of results generated by CPFlow.}
    \begin{ruledtabular}
    \renewcommand{\arraystretch}{1.5}
    \begin{tabular}{cccccc}
		\multirow{2}{*}{Target} & \multicolumn{3}{c}{State encoding}                                 & \multicolumn{2}{c}{Evolution}           \\
		                        & Diverging flow    & 2D vortex ($\psi_+$) & 2D vortex ($\psi_-$) & $t=\pi/4$         & $t=\pi/2$          \\ \hline
		$1-\mathrm{F}$   & $8\times 10^{-6}$ & $8\times 10^{-3}$    & $8\times 10^{-3}$    & $5\times 10^{-7}$ & $8\times 10^{-10}$
	\end{tabular}
    \label{table1}
    \end{ruledtabular}
\end{table}

\section{Experimental details}
\subsection{Device information}
\begin{figure*}
    \centering
    \includegraphics[width=\linewidth]{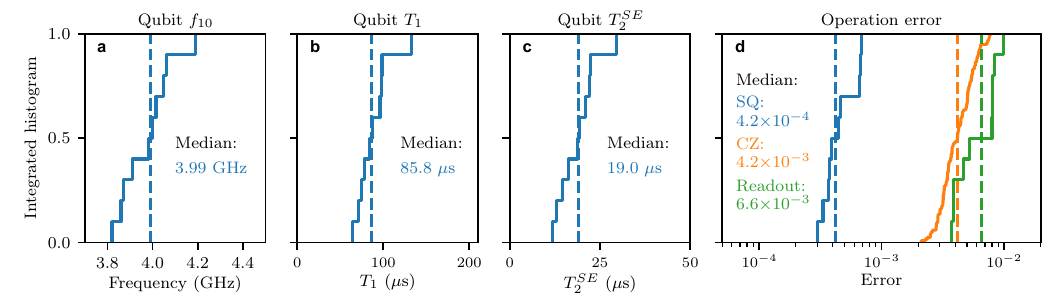}
    \caption{Integrated histograms of various device performance parameters. (a) Qubit idle frequency. (b) Qubit relaxation time measured at the idle frequency. (c) Qubit dephasing time measured using spin echo sequence. (d) Simultaneous operation errors of single-qubit gates (blue), two-qubit CZ gates (orange), and readout (green). Dashed lines indicate the median values. The errors of quantum gates are Pauli errors obtained through simultaneous cross-entropy benchmarking, and the readout error is calculated as $e_r=1-(f_0+f_1)/2$ with $f_{0(1)}$ denoting the measure fidelities of the state $|0\rangle$ ($|1\rangle$).}
    \label{fig:supp_fig_device_performance}
\end{figure*}

Our experiments are performed on a quantum processor with frequency-tunable transmon qubits encapsulated in a $11\times 11$ square lattice and tunable couplers between adjacent qubits, with the overall design similar to that in Ref.~\cite{SM_Xu2023_digital}. 
The maximum resonance frequencies of the qubit and coupler are around 4.5 GHz and 9.0 GHz, respectively. The effective coupling between two neighboring qubits can be dynamically activated by biasing the sandwiched coupler down close to qubits' interaction frequency. Each qubit capacitively couples to its own readout resonator at the frequency of around 6.5 GHz for dispersively readout. As illustrated in Fig.~\ref{fig:figure1}, ten qubits are used in our experiments, with the characterized properties summarized in Fig.~\ref{fig:supp_fig_device_performance}. 

\subsection{Circuit implementation}
We employ the native gate set \{$U(\theta, \varphi, \lambda)$, CZ\} to implement the desired experimental circuits, where $U(\theta,\varphi,\lambda)=R_z(\varphi)R_y(\theta)R_z(\lambda)$ denotes a generic single-qubit gate with three Euler angles, and CZ denotes the two-qubit CZ gate that fits the layout topology of our device. 
In practice, each generic single-qubit gate is achieved with a virtual Z rotation~\cite{SM_PhysRevA.96.022330} followed by a 30-ns long microwave pulse shaped with DRAG technique~\cite{SM_PhysRevLett.119.180511}, which can be formulated as $R_{\varphi+\pi/2}(\theta)R_z(\varphi+\lambda)$ where the subscript $\varphi+\pi/2$ refers to the angle of the equatorial rotation axis with respect to the $x$-axis. We realize two-qubit CZ gates by bringing $|11\rangle$ and $|02\rangle$ (or $|20\rangle$) of the qubit pairs in near resonance and activating the coupling for a specific time of 30 ns to achieve a complete cycle between $|11\rangle$ and $|02\rangle$ (or $|20\rangle$), with the calibration procedure detailed in Ref.~\cite{SM_quantum_adversarial_learning}.

\begin{figure*}
    \centering
    \includegraphics[scale=1]{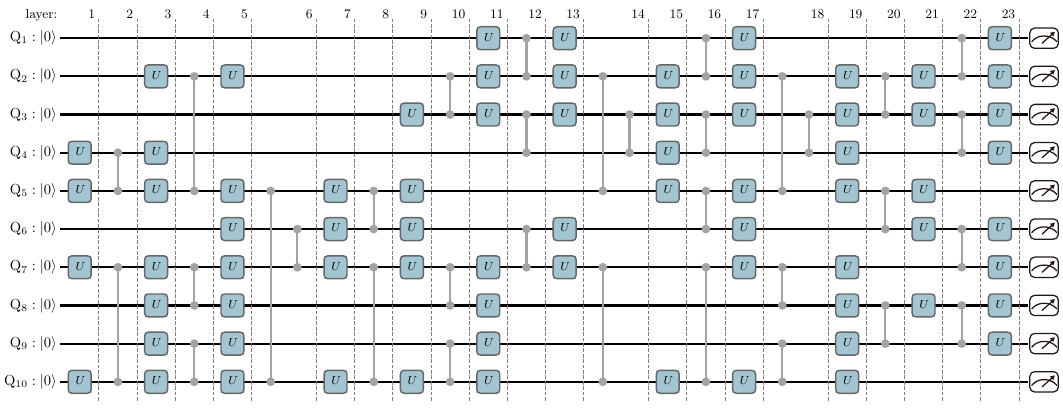}  
    \caption{Experimental circuit for encoding the spin-up component for the 2D vortex. Starting with the ground state $|0\rangle^{\otimes 10}$, we apply parallel single- and two-qubit gates layer by layer with the circuit depth up to 23. In practice, each layer of parallel single-qubit (two-qubit) gate requires a duration of 30 (40) ns. Hence, the total execution time is 810 ns, which is still much shorter than the median lifetime of qubits around 90 $\mu$s. At the end of the circuit, we measure all qubits simultaneously to reconstruct the density and momentum fields. Each round rectangle with label $U$ denotes a generic single-qubit gate.}
    \label{fig:supp_fig_circuit}
\end{figure*}

Each experimental circuit is aligned as a staggered arrangement of single- and two-qubit gate layers, with a typical experimental circuit shown in Fig.~\ref{fig:supp_fig_circuit} and all aligned circuit depths listed in Tab.~\ref{circuit_depth}. CZ gate parameters are optimized for each two-qubit gate layer. We perform simultaneous cross entropy benchmarking (XEB) \cite{SM_boixo_characterizing_2018, SM_Arute2019QuantumSupremacy} to characterize the performance of the quantum gates used in circuits with the statistical result shown in Fig.~\ref{fig:supp_fig_device_performance}(d).

\begin{table}[h]
    \centering
    \caption{Depths of aligned circuits.}
    \renewcommand{\arraystretch}{1.5}
    \setlength{\tabcolsep}{20pt}
    \begin{tabular}{cccccc}
        \hline\hline
	& $t=0$ & $t=\pi/4$ & $t=\pi/2$ \\
        \hline
	Diverging flow & 13 & 33 & 19 \\
	2D vortex ($\psi_+$) & 23 & 45 & 29 \\
	2D vortex ($\psi_-$) & 27 & 49 & 29 \\
        \hline\hline
    \end{tabular}
    \label{circuit_depth}
\end{table}

\subsection{Measurement of quantum state}
Here we derive the matrix elements of the measurement operator of density and momentum.
To facilitate this analysis, we introduce $\phi(m,l)=\psi_{2^{n_x}m+l}$, where the indices $l=0,1,\ldots,2^{n_x}-1$ and $m=0,1,\ldots,2^{n_y}-1$ denote the discrete coordinates in the $x$ and $y$ directions, respectively. 
Consequently, the density is measured as
\begin{equation}
    \rho(m,l) = \phi^*(m,l)\phi(m,l)
    = \psi^*_{2^{n_x}m+l}\psi_{2^{n_x}m+l}
    = \bra{\psi}\,\hat{\rho}^{(m,\,l)}\,\ket{\psi},
\end{equation}
and the matrix elements of the measurement operator are
\begin{equation}
    \hat{\rho}_{j,\,k}^{(m,\,l)} = \delta_{j,\,2^{n_x}m+l}\delta_{2^{n_x}m+l,\,k}.
\end{equation}
Note that the whole density field can be obtained from a single measurement.

Extracting momentum field is much more complicated. We employ the finite difference method to approximate derivatives in the momentum expression in Eq.~\eqref{eq:SM_rho_vel}. In practice, we replace non-bounded (bounded) derivatives with central (forward or backward) differences. The derivation for non-bounded momentum is provided below. 

The momentum is re-expressed as
\begin{align}
    \vec{J}(m,l)
    =\ & \frac{i}{2}\left[\phi(m,l)\bn\phi^*(m,l) - \phi^*(m,l)\bn\phi(m,l) \right]
    \notag \\
    \approx\ & \frac{i}{4\dx}\left[\left(\phi^*(m,l+1)-\phi^*(m,l-1)\right)\phi(m,l) - \phi^*(m,l)\left(\phi(m,l+1) - \phi(m,l-1) \right) \right]\vec{e}_x
    \notag \\
    &+ \frac{i}{4\dy}\left[\left(\phi^*(m+1,l)-\phi^*(m-1,l)\right)\phi(m,l) - \phi^*(m,l)\left(\phi(m+1,l) - \phi(m-1,l) \right) \right]\vec{e}_y
    \notag \\
    =\ & \frac{i}{4\dx}\left[\phi^*(m,l+1)\phi(m,l) - \phi^*(m,l-1)\phi(m,l) - \phi^*(m,l)\phi(m,l+1) + \phi^*(m,l)\phi(m,l-1) \right]\vec{e}_x
    \notag \\
    &+ \frac{i}{4\dy}\left[\phi^*(m+1,l)\phi(m,l) - \phi^*(m-1,l)\phi(m,l) - \phi^*(m,l)\phi(m+1,l) + \phi^*(m,l)\phi(m-1,l) \right]\vec{e}_y
    \notag \\
    =\ & \frac{i}{4\dx}\left(\psi_{2^{n_x}m+l+1}^*\psi_{2^{n_x}m+l} - \psi_{2^{n_x}m+l-1}^*\psi_{2^{n_x}m+l} - \psi_{2^{n_x}m+l}^*\psi_{2^{n_x}m+l+1} + \psi_{2^{n_x}m+l}^*\psi_{2^{n_x}m+l-1}\right)\vec{e}_x
    \notag \\
    &+ \frac{i}{4\dy}\left(\psi_{2^{n_x}(m+1)+l}^*\psi_{2^{n_x}m+l} - \psi_{2^{n_x}(m-1)+l}^*\psi_{2^{n_x}m+l} - \psi_{2^{n_x}m+l}^*\psi_{2^{n_x}(m+1)+l} + \psi_{2^{n_x}m+l}^*\psi_{2^{n_x}(m-1)+l}\right)\vec{e}_y
    \notag \\
    =\ & \bra{\psi}\,\hat{\vec{J}}^{(m,\,l)}\,\ket{\psi},
\end{align}
with matrix elements
\begin{align}
    \hat{\vec{J}}^{(m,\,l)}_{j,\,k}
    = \frac{i}{2}\bigg[&\delta_{2^{n_x}m+l,\,k}\bigg(\frac{\delta_{j,\,2^{n_x}m+l+1} - \delta_{j,\,2^{n_x}m+l-1}}{2\dx}\vec{e}_x + \frac{\delta_{j,\,2^{n_x}(m+1)+l}-\delta_{j,\,2^{n_x}(m-1)+l}}{2\dy}\vec{e}_y \bigg) 
    \notag \\
    &- \delta_{j,\,2^{n_x}m+l}\bigg(\frac{\delta_{2^{n_x}m+l+1,\,k} - \delta_{2^{n_x}m+l-1,\,k}}{2\dx}\vec{e}_x + \frac{\delta_{2^{n_x}(m+1)+l,\,k}-\delta_{2^{n_x}(m-1)+l,\,k}}{2\dy}\vec{e}_y \bigg) \bigg],
    \label{eq:SM_J_operator}
\end{align}
grid spacings $\dx=2\pi/2^{n_x}$ and $\dy=2\pi/2^{n_y}$, the Kronecker delta $\delta$, and Cartesian unit vectors $\{\vec{e}_x,\vec{e}_y\}$. 

For vortical flows, we measure $\rho_\pm(m,l)$ and $\vec{J}_{\pm}(m,l)$ using the same method, and then compute $\vec{u} = (\vec{J}_+ + \vec{J}_-)/(\rho_+ + \rho_-)$ and $\omega=\p u_y/\p x - \p u_x/\p y$. 

\begin{figure*}
    \centering
    \includegraphics[scale=1]{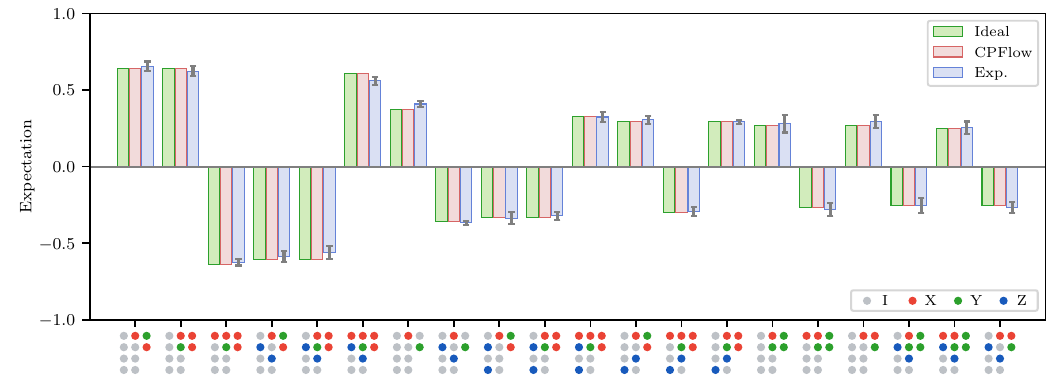}
    \caption{Expectation values of decomposed Pauli strings for the initial state of the diverging flow. 
    We plot the first 20 expectation values obtained from the exact solution (green), CPFlow simulation (red), and experiment (blue) in decreasing order according to the absolute values. Error bars on experimental data represent ten standard deviations. Each layout pattern under $x$-axis denote a Pauli string with each qubit colored according to the corresponding Pauli operator.}
    \label{fig:supp_fig_expectation}
\end{figure*}

In principle, each generic $n$-qubit observable $O$ can be decomposed as a sum of Pauli strings
\begin{equation}\label{eq:SM_decompose_J}
    O = \sum_{j=0}^{4^n-1} \frac{\text{Tr}(OP_j)}{2^n}P_j, \quad P_j\in\{\sigma^{\otimes n}\}
\end{equation}
which requires an exponentially increasing number of measurements. 
However, for the momentum operator $\hat{\vec{J}}^{(m,\,l)}$, only a few terms in Eq.~\eqref{eq:SM_decompose_J} have non-zero coefficients, which significantly reduce the experimental runtime. 
As a result, 5120 Pauli strings are required to reconstruct the momentum field. Furthermore, since some local Pauli strings, e.g., ZIZZIXXXYX, can be calculated from the measured probability distribution of certain global Pauli strings, e.g., ZZZZZXXXYX, the measurement number is eventually reduced to 62. Here we plot the first 20 Pauli string expectation values in decreasing order for the diverging flow case in Fig.~\ref{fig:supp_fig_expectation}, which shows decent agreement with the exact solution.

\begin{figure*}
    \centering
    \includegraphics[scale=1]{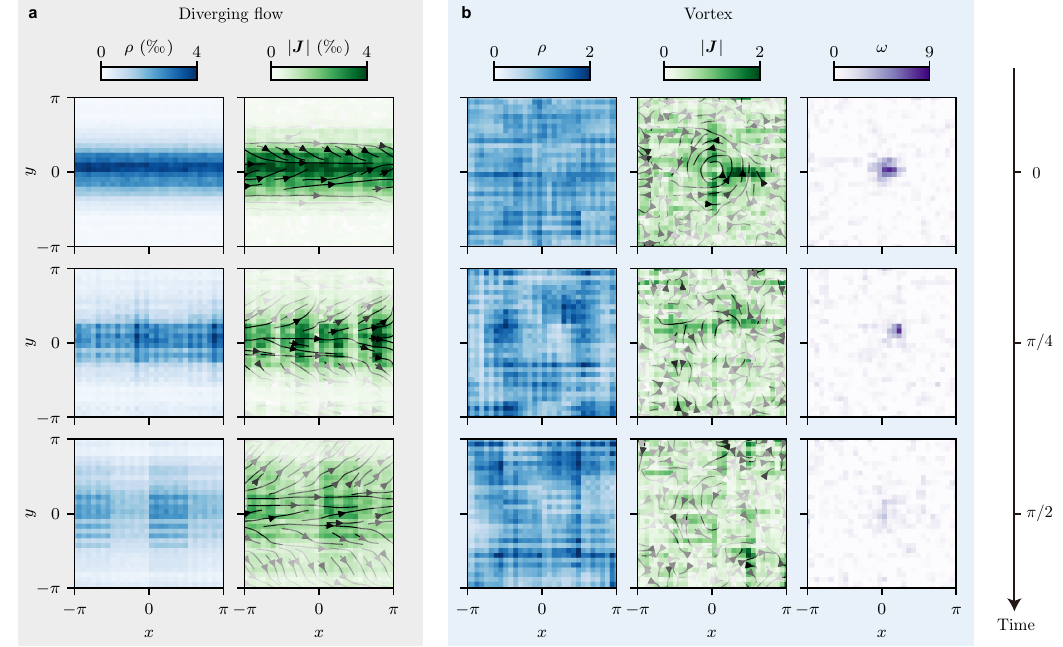}
    \caption{Experimental results on the temporal evolution (from top to bottom) of flow fields. 
    (a) Density (left) and momentum (right) contours with streamlines for the 2D diverging flow. 
    (b) Density (left), momentum (middle), and vorticity (right) contours for the 2D vortex.}
    \label{fig:supp_fig_exp_field}
\end{figure*}

In summary, we perform 63 measurements to reconstruct the density field and momentum field for each circuit. Each measurement involves $10^5$ single shots to build the probability distribution, consuming approximately 20 s at a sampling rate of 5 kHz. The experiment is repeated five times for each flow case. The average results for all flow cases in main text are shown in Fig.~\ref{fig:supp_fig_exp_field}. 

\section{Effects of quantum errors}
Although the experimental results on the superconducting processor capture the major features of fluid dynamics in all flow cases, there are some notable artifacts, such as regular stripe patterns in the reconstructed flow fields in Fig.~2(c). Here we investigate the origins of these artifacts with numerical simulations.

\subsection{Stripe-like artifacts}
\begin{figure*}
    \centering
    \includegraphics[scale=1]{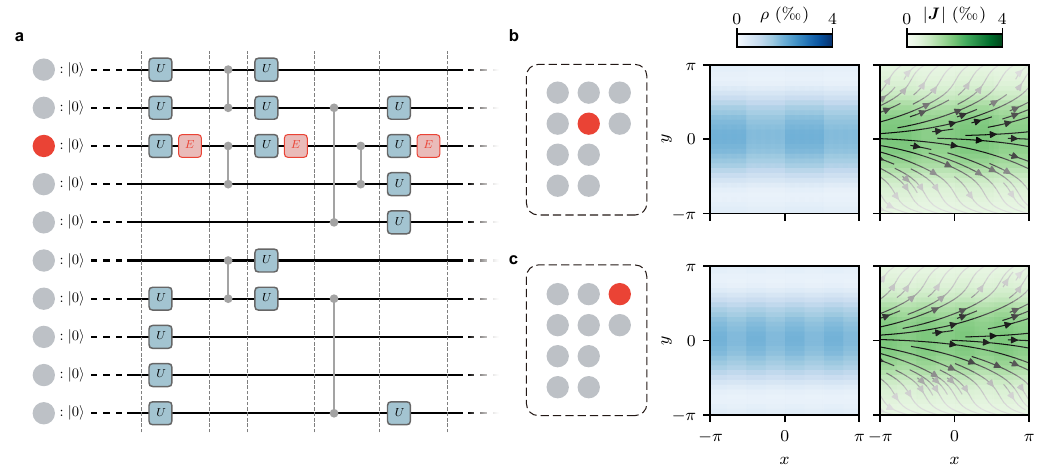}
    \caption{Numerical simulations for showing the origin of stripe-like artifacts in the experimental results for the 2D diverging flow. 
    (a) We apply an extra error gate after each single-qubit gate on a certain qubit as a local perturbation. The error gate is selected as $R_x(0.025)$ with an equivalent error rate of around $5 \times10^{-4}$. (b,c) Simulation results for the evolution of the 2D diverging flow at $t=\pi/2$ with two different choices of the error qubit as shown in dashed rectangles.}
    \label{fig:supp_fig_stripe_pattern}
\end{figure*}

The ability of the finite number of qubits to encode an exponentially large fluid state suggests that local perturbations can exert a global influence on the encoded flow field. 
As illustrated in Fig.~\ref{fig:supp_fig_stripe_pattern}(a), we demonstrate this phenomenon by applying an extra error gate after each single-qubit gate on a certain qubit. The numerical simulation results exhibit different regular stripe patterns depending on the choice of the error qubit in Figs.~\ref{fig:supp_fig_stripe_pattern}(b) and (c).

\subsection{Deviation of $J_y$ in the diverging flow}
\begin{figure*}
    \centering
    \includegraphics[scale=1]{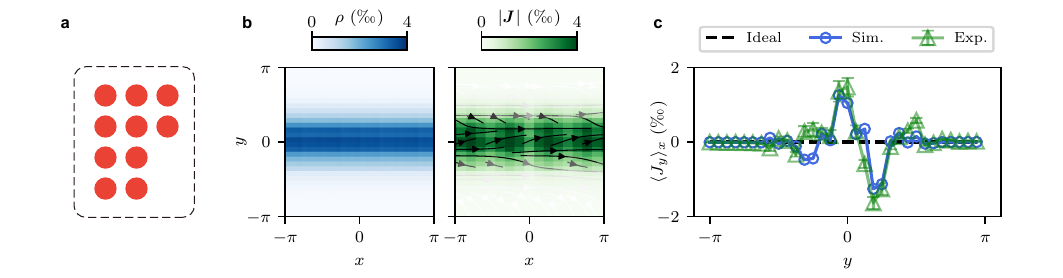}
    \caption{Numerical simulations for showing the origin of the notable deviation of $J_y$ in the experimental results for the 2D diverging flow.
    (a) Similar to Fig.~\ref{fig:supp_fig_stripe_pattern}, all qubits in red indicate that each single-qubit gate in the circuit is appended with an error gate. Error gates are selected as $U(\theta,\phi,\lambda)$ with parameters randomly sampled from uniform distribution of $[-0.045, 0.045]$. The average equivalent error rates are around $5 \times10^{-4}$. To obtain the desired $x$-averaged profile of $J_y$, we have carefully searched the random seed. 
    (b) Reconstructed density (left) and momentum (right) contours from the numerical simulation. 
    (c) Comparison of the $x$-averaged profile of $J_y$ among results from the exact solution (black), simulation (blue), and experiment (green).}
    \label{fig:supp_fig_deviation_of_Jy}
\end{figure*}

We generate a circuit for encoding the 2D diverging flow by applying an extra error gate after each single-qubit gate on all qubits, which yields a similar $x$-averaged profile of $J_y$ with the experimental result in Fig.~\ref{fig:supp_fig_deviation_of_Jy}. 

\section{Complexity analysis}
We analyze the algorithm complexities of simulating fluid flows via Hamiltonian simulation.
Consider a $D$-dimensional flow using $nD$ qubits with $N=2^n$ grid points in each dimension and total number $N^D=2^{nD}$ of grid points. 

First, we address the time complexity of state preparation. 
The preparation of an initial state involves breaking down each unitary transformation on $nD$ qubits.  
Various quantum algorithms and their complexities for state preparation are summarized in Ref.~\cite{SM_Shao2018_Quantum}. 
Currently, no quantum algorithm achieves simultaneous logarithmic time complexity in terms of $N^D$, $\kappa$, and $1/\varepsilon$,  
where $\kappa$ denotes the condition number of the state vector, and $\varepsilon$ characterizes accuracy. 
In general, preparing any $nD$-qubit initial state needs a prohibitive circuit depth of $\mathcal{O}(2^{nD})$, which challenges the quantum advantage. 
This problem can be addressed through the implementation of quantum random access memory~\cite{SM_Giovannetti2008_Quantum}. 
Additionally, variational quantum algorithms can facilitate generating the approximate quantum circuit for a given initial velocity with a polynomial scale.

Second, we analyze the time complexity of state evolution. 
The quantum gate complexity per temporal increment is $ S_q = \mathcal{O}(n^2D) $~\cite{SM_Nielsen2010_Quantum}. 
By contrast, the corresponding complexity using the fast Fourier transform is $ S_c = \mathcal{O}(nD2^{nD})$ in classical computing. 
For the state evolution, quantum computing demonstrates a remarkable speedup, with the ratio $R_q = \mathcal{O}(2^{nD}/n)$ compared to its classical counterpart. 

Third, we analyze the measurement complexity.
To reconstruct the entire flow field, the expectation of $\mathcal{O}(4^{nD})$ Pauli strings must be measured.
Each measurement involves $\Od{100\times 2^{nD}}$ single shots to build the probability distribution and is repeated for several times.
This requirement results in high time complexity for the measurement. 
Therefore, it is imperative to devise measurement operators that either restrict flow statistics to a minimal set of qubits or allow an expansion into polynomial-scale Pauli strings. 

%

\end{document}